\documentclass[prl,showpacs,superscriptaddress,nofootinbib,10pt]{revtex4-2}
\usepackage{color,amsmath,amssymb,graphicx,latexsym,lineno}
\usepackage{threeparttable,multirow,txfonts,subfigure,booktabs}
\usepackage{adjustbox}
\usepackage{soul}
\usepackage{ulem}

\newcommand{\rev}[1]{{#1}}
\newcommand{\wm}[1]{{#1}}

\graphicspath{{./}{figure_for_paper/}}

\begin{document}

\title{Transient Large-Scale Anisotropy in TeV Cosmic Rays due to an Interplanetary Coronal Mass Ejection}

 
\author{Zhen Cao}
\affiliation{State Key Laboratory of Particle Astrophysics \& Experimental Physics Division \& Computing Center, Institute of High Energy Physics, Chinese Academy of Sciences, 100049 Beijing, China}
\affiliation{University of Chinese Academy of Sciences, 100049 Beijing, China}
\affiliation{TIANFU Cosmic Ray Research Center, Chengdu, Sichuan,  China}
 
\author{F. Aharonian}
\affiliation{TIANFU Cosmic Ray Research Center, Chengdu, Sichuan,  China}
\affiliation{University of Science and Technology of China, 230026 Hefei, Anhui, China}
\affiliation{Yerevan State University, 1 Alek Manukyan Street, Yerevan 0025, Armenia}
\affiliation{Max-Planck-Institut for Nuclear Physics, P.O. Box 103980, 69029  Heidelberg, Germany}
 
\author{Y.X. Bai}
\affiliation{State Key Laboratory of Particle Astrophysics \& Experimental Physics Division \& Computing Center, Institute of High Energy Physics, Chinese Academy of Sciences, 100049 Beijing, China}
\affiliation{TIANFU Cosmic Ray Research Center, Chengdu, Sichuan,  China}
 
\author{Y.W. Bao}
\affiliation{Tsung-Dao Lee Institute \& School of Physics and Astronomy, Shanghai Jiao Tong University, 200240 Shanghai, China}
 
\author{D. Bastieri}
\affiliation{Center for Astrophysics, Guangzhou University, 510006 Guangzhou, Guangdong, China}
 
\author{X.J. Bi}
\affiliation{State Key Laboratory of Particle Astrophysics \& Experimental Physics Division \& Computing Center, Institute of High Energy Physics, Chinese Academy of Sciences, 100049 Beijing, China}
\affiliation{University of Chinese Academy of Sciences, 100049 Beijing, China}
\affiliation{TIANFU Cosmic Ray Research Center, Chengdu, Sichuan,  China}
 
\author{Y.J. Bi}
\affiliation{State Key Laboratory of Particle Astrophysics \& Experimental Physics Division \& Computing Center, Institute of High Energy Physics, Chinese Academy of Sciences, 100049 Beijing, China}
\affiliation{TIANFU Cosmic Ray Research Center, Chengdu, Sichuan,  China}
 
\author{W. Bian}
\affiliation{Tsung-Dao Lee Institute \& School of Physics and Astronomy, Shanghai Jiao Tong University, 200240 Shanghai, China}
 
\author{A.V. Bukevich}
\affiliation{Institute for Nuclear Research of Russian Academy of Sciences, 117312 Moscow, Russia}
 
\author{C.M. Cai}
\affiliation{School of Physical Science and Technology \&  School of Information Science and Technology, Southwest Jiaotong University, 610031 Chengdu, Sichuan, China}
 
% ...
\collaboration{The LHAASO Collaboration}

\email[E-mail: ]{david.ruf@mahidol.ac.th, xaye@pmo.ac.cn, warit.mit@mahidol.ac.th, zhangyi@pmo.ac.cn}

\date{\today}

\begin{abstract}
Large- or medium-scale cosmic ray anisotropy at TeV energies has not previously been confirmed to vary with time.  Transient anisotropy changes have been observed 
below 150 GeV, especially near the passage of an interplanetary shock and coronal mass ejection containing a magnetic flux rope ejected by a solar storm, which can trigger a geomagnetic storm with practical consequences.  In such events, cosmic rays provide remote sensing of the magnetic field properties.  Here we report the observation of transient large-scale anisotropy in TeV cosmic ray ions using data from the Large High Altitude Air Shower Observatory (LHAASO). We analyze hourly skymaps of the transient cosmic ray intensity excess or deficit, the gradient of which indicates the direction and magnitude of transient large-scale anisotropy across the field of view.  We observe enhanced anisotropy above typical hourly fluctuations with $>$5$\sigma$ significance during some hours of November 4, 2021, in separate data sets for four primary cosmic ray energy ranges of median energy from $E$=0.7 to 3.1 TeV.  The gradient varies with energy as $E^{\gamma}$, where $\gamma\approx-0.5$.  At a median energy $\leq$1.0 TeV, this gradient corresponds to a dipole anisotropy of at least 1\%, or possibly a weaker anisotropy of higher order.  This new type of observation opens the opportunity to study interplanetary magnetic structures using air shower arrays around the world, complementing existing {\it in situ} and remote measurements of plasma properties.
\end{abstract}


\maketitle
{\it Introduction.}--- The flux and anisotropy of Galactic cosmic ray (CR) ions are generally considered to be constant over decades before approaching our heliosphere. Within the heliosphere, various temporal variations due to solar and heliospheric phenomena have been documented for GeV-range ions. These include the effects of the $\approx$11-year solar activity (sunspot) cycle \cite{11cycle}, the $\approx$22-year solar magnetic cycle \cite{22cycle,JokipiiThomas81}, the $\approx$27-day solar rotation \cite{27cycle}, and transient effects such as Forbush decreases caused by solar storms and solar wind variations \cite{Forbush1937}. However, the maximum cosmic ray energy affected by these phenomena remains unclear. 

As solar and heliospheric effects diminish with increasing particle energy, varying by at most several percent above approximately 20 GeV \cite{ruffolo2020time}, measuring these effects at much higher energies requires the use of large ground-based or underground detectors.
Underground muon detectors have detected transient Forbush decreases in GCR ion (mostly proton) intensity up to a median rigidity (i.e., momentum per charge in GV, which for high-energy protons is very similar to the energy in GeV)
of about 331 GV \citep{Sakakibara1979,Munakata2000}, and clear precursory large-scale anisotropy before the arrival of a CME-driven interplanetary shock for a median rigidity up to 145 GV \citep{Munakata2000}.
A recent analysis of ground-based neutron monitor and muon detector data during the Forbush decrease of November 3-5, 2021 reported a strong ($>$1\%) anisotropy signal up to $\sim$65 GV with a rigidity power-law index between +0.5 and -1.5  \wm{($\gamma_1$ in Equation~(3) of \citep{Munakata2022})}, which indicates that the anisotropy could possibly extend to much higher rigidity.

There have been very few reports of solar and heliospheric effects on \rev{TeV} cosmic rays.
Time variability has been noted in the Sun shadow of \rev{TeV} ions, including dependence on the interplanetary magnetic field \rev{\cite{IMFonCRshadow,LHAASOsun24}}, variation with the solar activity cycle \cite{solarcycleonCR,Chen17}, and the combined effects of solar storms \cite{Tibet2018}.
However, the large- or medium-scale cosmic ray anisotropy at the TeV range or higher energies is usually reported to be constant in time.
An exception is a report that the ``loss-cone'' deficit in cosmic ray anisotropy \citep{Nagashima98} 
as observed by the Milagro observatory at a median energy of 6 TeV became stronger with time during 2000–2007 \citep{Abdo09}.
In contrast, the Tibet air-shower experiment reported no such temporal changes throughout 1999-2008 for median energies of 4-11 TeV \citep{Amenomori12}, nor did the ARGO-YBJ experiment over 2008-2012 at a median energy of 7 TeV \citep{Bartoli18}, and in the southern sky, the IceCube Collaboration reported no significant changes over 2009-2015 \citep{Aartsen16}.
At slightly lower energy, solar cycle effects on the yearly large-scale cosmic ray anisotropy 
have been reported using the Matsushiro underground muon detector, which is sensitive to a median CR rigidity of 0.6 TeV\cite{munakata2010solar}.

In this study, we have developed techniques to analyze hourly anisotropy data from air shower array data, specifically from
the Large High-Altitude Air Shower Observatory (LHAASO). 
We present the first observation of transient variation in large-scale anisotropy at energies both below and above 1 TeV, associated with passage of a shock and interplanetary coronal mass ejection (ICME) on November 4, 2021.

\begin{table*}[t]
\centering
\caption{Detection of transient large-scale CR anisotropy on November 4, 2021.}
\begin{tabular}{c c c c c c c c c}
\hline
& \textrm{\rev{$N_{hit} $ or $N_{filtE}$}} & \textrm{Median primary} & & \textrm{Gradient} & \textrm{Formal} & \textrm{Normalized} & \textrm{Chance} & \textrm{Time of max.} \\
\textrm{Detector} & \textrm{\rev{range}} & \textrm{CR energy (TeV)} & \textrm{Time (UT)} & \textrm{magnitude, $g$ (\%)} & \textrm{significance ($\sigma$)} & \textrm{significance ($\sigma$)} & \textrm{probability, $p$} & \textrm{significance?} \\
\hline
WCDA& $[30,40)$   & 0.7 & 10:00 - 11:00 & 1.20  & 10.98 & 7.06 & 1.67e-12 & \checkmark \\
    & $[40,60)$   & 1.0 & 10:00 - 11:00 & 1.11  & 11.46 & 7.42 & 1.17e-13 & \checkmark \\
    & $[60,100)$  & 1.5 & 10:00 - 11:00 & 0.78 & 7.17 & 5.11 & 3.22e-07 & \\
    & $[60,100)$  & 1.5 & 13:00 - 14:00 & 0.92 & 8.48 & 6.08 & 1.20e-09 & \checkmark \\
    & $[100,320)$ & 3.1 & 10:00 - 11:00 & 0.56 & 4.80 & 3.41 & 6.50e-04 & \\
    & $[100,320)$ & 3.1 & 13:00 - 14:00 & 0.81 & 7.11 & 5.14 & 2.75e-07 & \checkmark  \\
    & $[320,2025)$& 12.6 & 10:00 - 11:00 & 0.52 & 1.84 & 1.53 & 1.26e-01 & \\
    & $[320,2025)$& 12.6 & 14:00 - 15:00 & 0.63 & 2.35 & 1.98 & 4.77e-02 & \checkmark \\
\hline
KM2A& $(10,+\infty)$ & 26.5 & 10:00 - 11:00 & 0.51 & 3.44 & 2.55 & 1.08e-02 & \\
    & $(10,+\infty)$ & 26.5 & 14:00 - 15:00 & 0.65 & 4.56& 3.42 & 6.26e-04 & \checkmark \\
\hline
\end{tabular}

\label{tab:list}
\end{table*}

{\it Measurement and Analysis Techniques.}---LHAASO is a large-scale dual-purpose facility for researching cosmic and gamma rays, at an altitude of 4410 meters in Sichuan Province, China. For cosmic-ray detection, its detectors cover over four orders of magnitude in energy. 
The 78,000-m$^2$ Water Cherenkov Detector Array (WCDA) is subdivided into 3120 units that detect CRs from below 1 TeV to several tens of TeV. 
The Kilometer Squared Array (KM2A), with 5216 surface scintillator detectors and 1188 underground muon detectors spread uniformly over 1.3 km$^2$, targets very-high-energy cosmic rays from tens of TeV to beyond 10 PeV. The Wide Field-of-view Cherenkov Telescope Array (WFCTA) offers high sensitivity to cosmic rays above 1 PeV, 
which is not required in the present work on \rev{TeV} CR anisotropy. 
The huge collection area of LHAASO detector arrays provides sufficient statistical accuracy for our measurement of percent-level anisotropy hour by hour. 

We select events with zenith angle $<$45$^\circ$ and \wm{only} hours with 100-120 million events for WCDA or 7-10 million events for KM2A. \rev{The shower energy is characterized by $N_{hit}$ for WCDA and by $N_{filtE}$ for KM2A. Both $N_{hit}$ and $N_{filtE}$ represent the number of triggered detectors after applying certain cut conditions (see Supplemental Material for details).} The median energy for each \rev{$N_{hit}$ or $N_{filtE}$} range is listed in Table \ref{tab:list}.  
Calculated distributions of primary cosmic ray energy for each range can be found in the Supplemental Material.


In our analysis, for each hour of data we create a skymap in the horizontal coordinate system, divided into segments of equal solid angle for 8 ranges in zenith angle \(\theta\) up to $\theta=45^\circ$ and 16 ranges in azimuth angle \(\phi\).  
The count in each segment is normalized by the average count for that $\theta$ range. 
The relative intensity \rev{in} each segment is then divided by the average background value for the same hour of day over \rev{two} preceding days and \rev{two} following days.  
\rev{Deviation of this relative intensity from 1 indicates a transient intensity excess or deficit.
}
We compare with values from the same direction and same hour of day so that, subject to Earth's rotation, the viewing direction is oriented toward a similar region of the sky. 
Note also that the sidereal anisotropy pattern and the Compton-Getting effect due to Earth's orbital motion combine in different ways at different times of year, so we only compare with background values \rev{on} 2 days \rev{before and after} the time of interest. \rev{To ensure an unbiased background estimate, data collected between November 2 and 6 were omitted to avoid contamination from the signal region.} 
In the Supplemental Material, we provide further details of this analysis technique, as well as a related technique that estimates the absolute skymap gradient for a given hour, which produces results consistent with known sidereal and Compton-Getting effects during background time periods, and after background subtraction yields similar gradient values during the event of interest.

We then define a skymap gradient \( \mathbf{g} \) to characterize the anisotropy of the relative intensity within the field of view (FOV) for a given time hour.
Each skymap comprises 128 segments, and we specify the direction cosines \rev{$x_k=\cos\theta_k\cos\phi_k$ and $y_k=\cos\theta_k\sin\phi_k$} for \rev{each} directional segment $k$. 
We model the relative intensity as \( 1 + g_x x_k + g_y y_k \) with least-squares fitting to derive the gradient components \( g_x \) and \( g_y \) in the East and North directions, respectively, and the gradient magnitude
$g = \sqrt{g_x^2 + g_y^2}$.


The uncertainty of each component $g_i$ can be defined in two ways.  
One is a formal fitting uncertainty $\sigma_f$ and the other is an empirical uncertainty $\sigma_g$.  
Distributions of formal significance $g_i/\sigma_f$ during quiet times without ICME effects are consistent with Gaussian distributions of mean zero and \rev{standard deviation} $w$ ranging from 1.0 at high energy to 1.6 at low energy. 
Thus the overall spread in $g_i$, including statistical and other effects, can be described by $\sigma_g=w\sigma_f$, and we derive a normalized significance $S_i=g_i/\sigma_g$ \rev{above typical hourly fluctuations}, which is more conservative (smaller) than the formal significance based on fitting uncertainty alone.
The probability density function (PDF) of the gradient magnitude 
during quiet times is the Rayleigh distribution
\rev{(see Supplemental Material).}
The \( p\text{-value} \), i.e., chance probability of obtaining a gradient magnitude \( g \), can be calculated from this distribution.
We then determine $S_g$, the normalized significance of $g$, that corresponds to a two-sided chance probability equal to that \( p\text{-value} \) for a Gaussian distribution, given the uncertainty $\sigma_g$, as well $S_f$, the formal significance based on $\sigma_f$.

\begin{figure*}[t]
\includegraphics[scale=0.5]{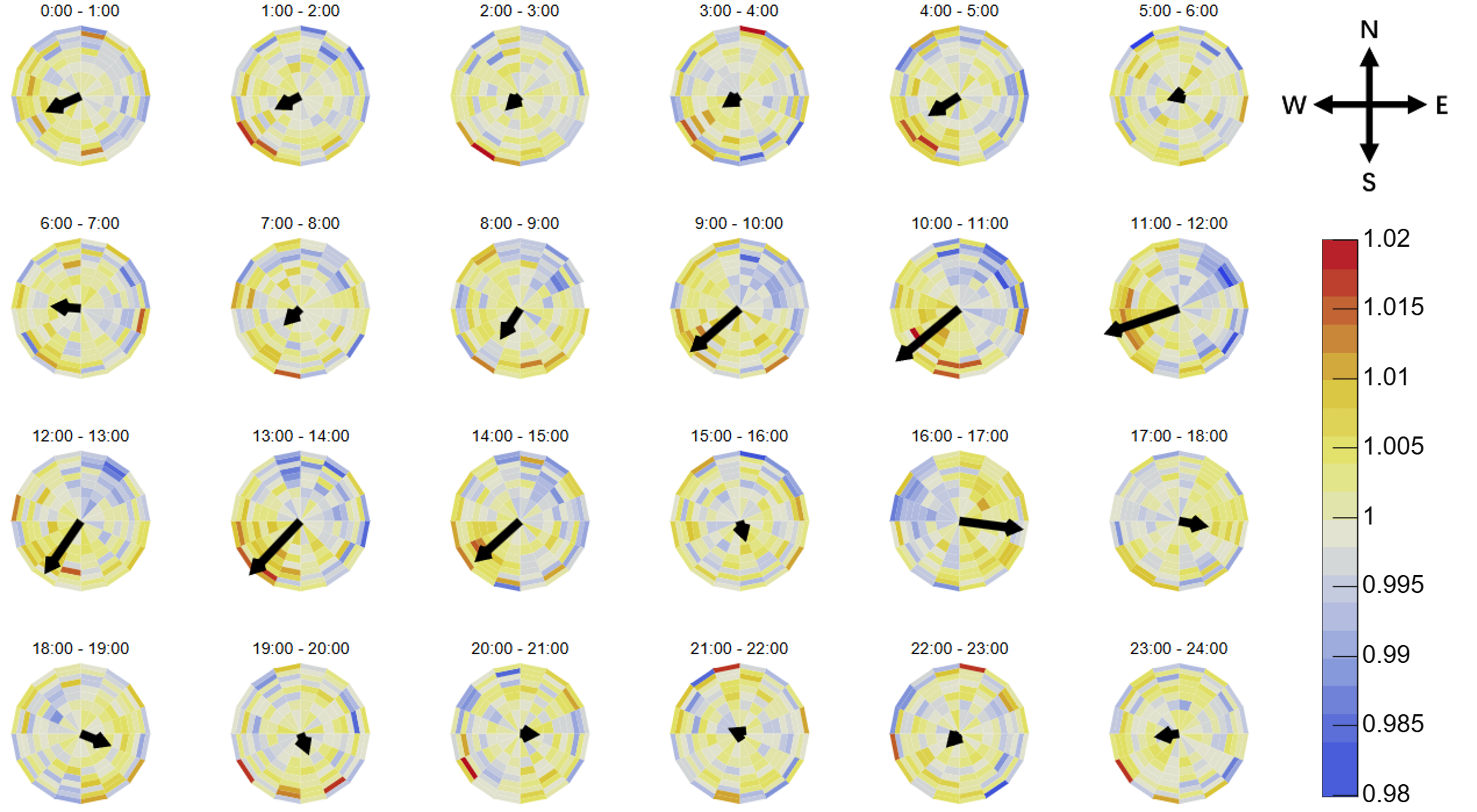}
\caption{Skymaps of relative CR intensity centered at the zenith and extending to zenith angle 45° (outer circle), for showers with $30 \leq N_{hit} < 40$ in LHAASO/WCDA for each hour in Universal Time (UT) on November 4, 2021.  
An Interplanetary Coronal Mass Ejection (ICME) impacted Earth at about 12:00 UT. Black arrow represents the best-fit CR gradient vector, with arrow length proportional to gradient magnitude; it points away from deficit areas (blue) and towards areas of enhancement (red). 
The anisotropy increased strongly a few hours before 1200 UT, the time of ICME arrival, with a maximum gradient magnitude of 1.2\% during 10:00-11:00 UT,  and continued to be strong for a few hours into ICME passage.}
    \label{fig:WCDA_nhit30-40_24hmap}
\end{figure*}

\begin{figure*}[t]
\includegraphics[scale=0.9]{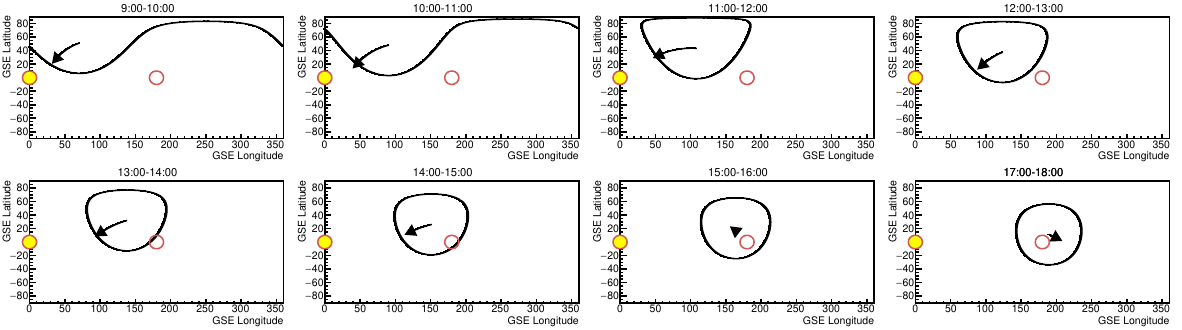}
\caption{Gradient of relative anisotropy (arrow with length proportional to gradient magnitude) and LHAASO field of view (FOV, within the curve) in GSE coordinates for each hour from 09:00 UT to 17:00 UT on November 4, 2021 for $30\leq N_{hit}<40$ (median energy 0.7 TeV).
Sunward (yellow symbol) and anti-Sunward (red circle) viewing directions are also indicated.   
Other than 15:00-16:00, when the FOV may have included a direction of minimal flux, the gradient has $\geq4\sigma$ significance for each hour and usually indicates a higher cosmic ray flux from directions closer to the Sunward viewing direction.
}
    \label{fig:WCDA_nhit30-40_GSE}
\end{figure*}

\begin{figure*}[t]
\includegraphics[scale=0.55]{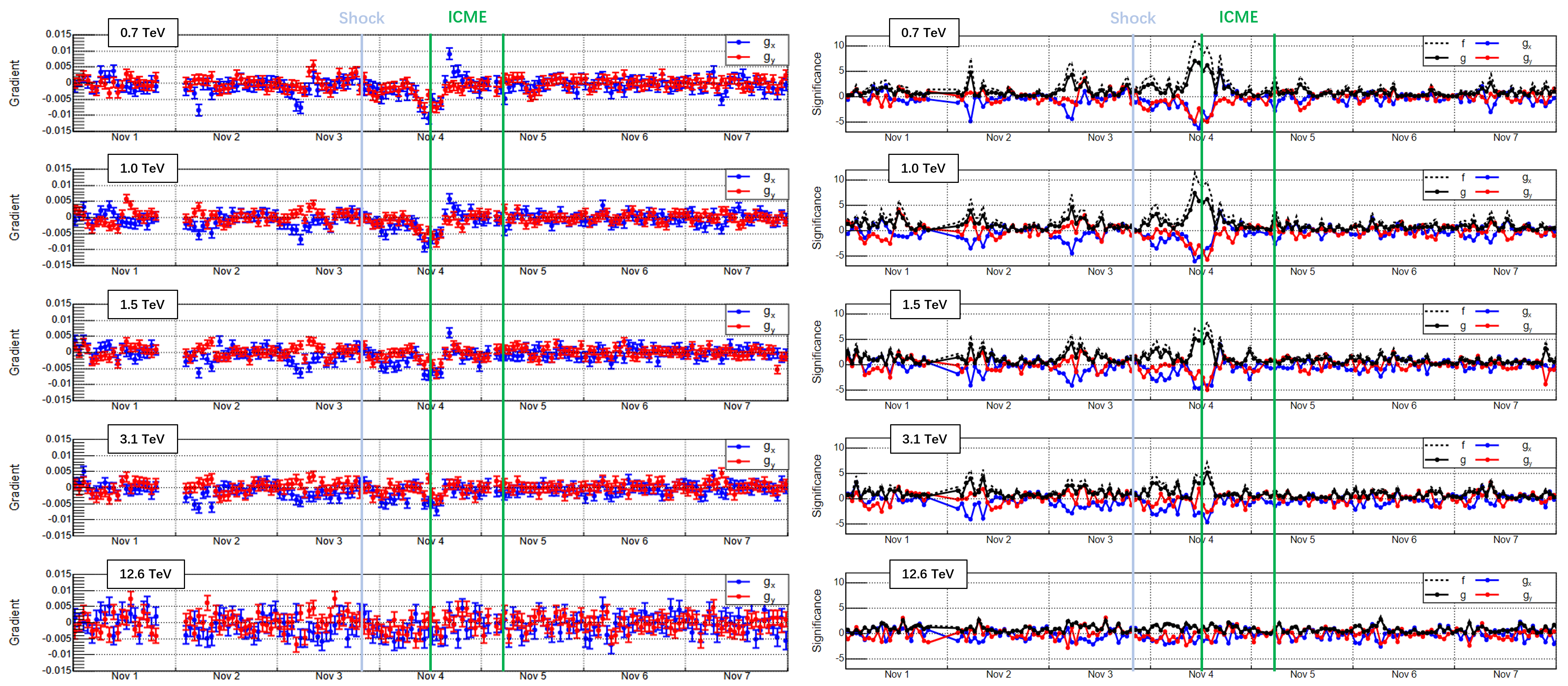}
\caption{Time series of skymap gradient components and magnitude and their significance for various WCDA energy ranges during November 1-7, 2021. Green lines indicate start and end times of ICME passage. Median primary CR energy is indicated at the top left of each panel.
Solid traces indicate normalized significance of the gradient magnitude and components, while the dashed trace indicates the formal significance of the magnitude based on fitting uncertainty alone.}
    \label{fig:WCDA_grad&sig}
\end{figure*}

\begin{figure}[t]
\includegraphics[scale=0.3]{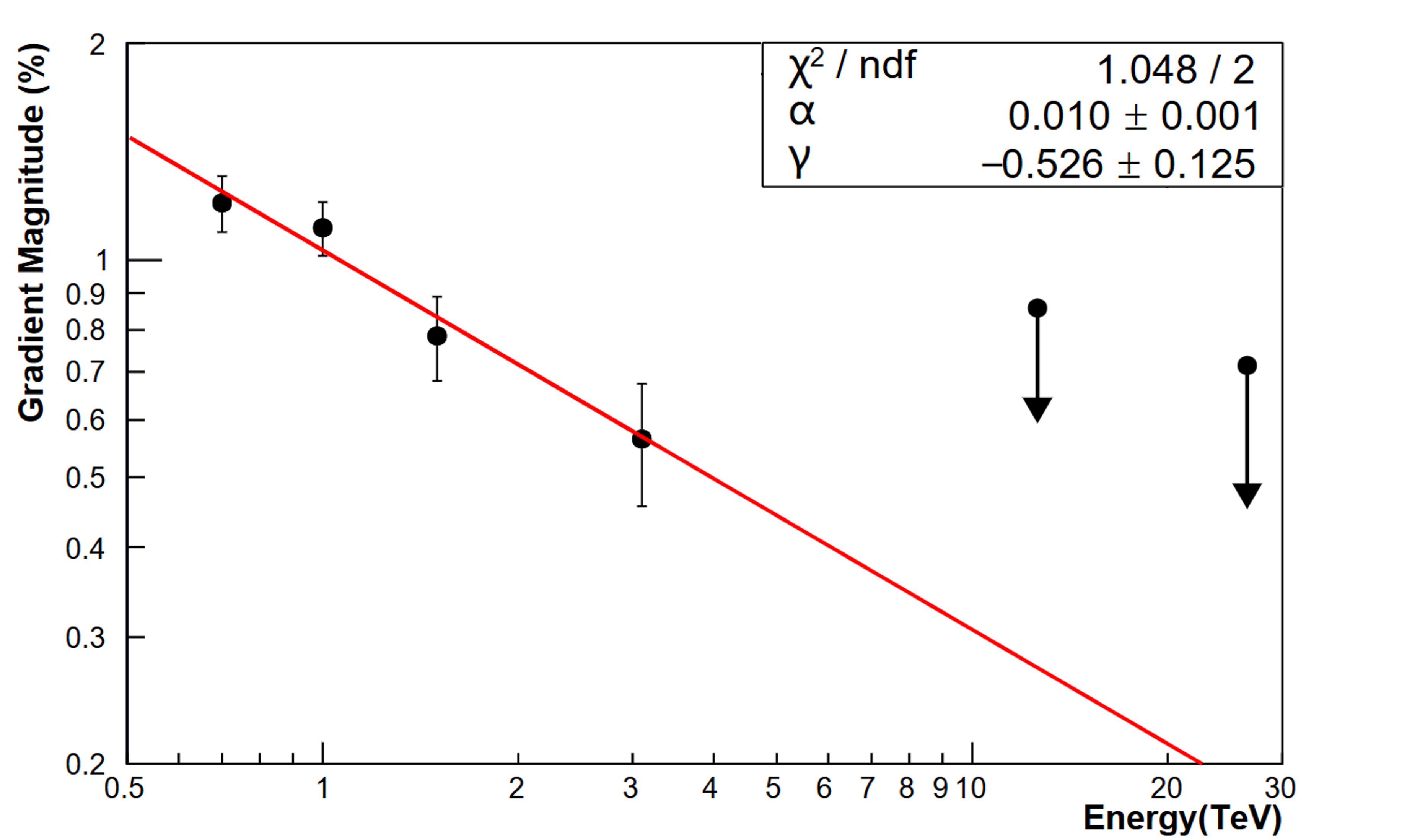}
\caption{Observed gradient magnitude $g$ vs.\ median CR energy during 10:00–11:00 UT on November 4, 2021, with power-law fit to the first four energy ranges.}
    \label{fig:WCDA_KM2A_amplitude}
\end{figure}


{\it Results.}--- We examined the ICME passage on November 4, 2021, for which a strong anisotropy in the GCR distribution has been reported at $\sim$65 GV and lower rigidities \citep{Munakata2022,Benella2025}. 
Figure \ref{fig:WCDA_nhit30-40_24hmap} presents skymaps for each hour in Universal Time (UT) on November 4, utilizing WCDA data with $N_{hit}$ values between 30 and 40. Each skymap is labeled with its corresponding hour. The color of each bin represents the relative intensity, and arrows illustrate the gradient $\mathbf{g}$, indicating the direction of higher relative intensity. The length of each arrow corresponds to the gradient's magnitude. Notably, between 09:00 and 15:00 UT, the gradient magnitude was much stronger than during quiet times. 

Figure \ref{fig:WCDA_nhit30-40_GSE} depicts the gradient directionality in Geocentric Solar Ecliptic (GSE) coordinates. The yellow circle designates the Sun's direction, while the hollow circle marks the \rev{anti-Sunward} direction. The solid black line outlines the FOV for that hour, with arrows representing the gradient direction, specific to the central time of that hour.  
\rev{The gradient pointed} towards the Sun's direction starting at 09:00 UT, 
\rev{shortly before the ICME arrival} at about 12:00 UT.
\rev{It weakened} significantly when the FOV \rev{was centered near the anti-Sunward} direction at 15:00-16:00 UT, and \rev{shifted} direction again at 16:00-17:00 UT. 
\rev{This is consistent with a transient deficit in the cosmic ray intensity near the anti-Sunward direction, resulting in a gradient away from that direction.}

Figure \ref{fig:WCDA_grad&sig} presents the gradients and significance values for five independent WCDA energy bins, with the median energy indicated in the top left corner of each panel. The light blue line denotes the shock arrival time, whereas the green lines mark the period of ICME passage. Note that the first four energy ranges all exhibit similar gradients in a similar direction during times near 12:00 UT.  A strong peak in significance is observed for 10:00–11:00 UT in the lower three $N_{hit}$ ranges. However, for $N_{hit}$ ranges of [60,100) and [100,320), 
the hour with the highest significance is 13:00–14:00 UT. 
Table \ref{tab:list} provides the gradient, significance, and chance probability values for various time intervals across different energy bands, including an indication of whether this was the time of maximum $g$ for each energy band. 
Notably, the first four energy bands exhibit $g$ values with normalized significance of $5\sigma$ or more.

Overall, the 10:00–11:00 UT interval shows the highest significance across all energy ranges. We specifically examine the gradient magnitude during this interval and fit it to a power-law spectrum $\alpha (E/\rm{TeV})^{\gamma}$ (Figure \ref{fig:WCDA_KM2A_amplitude}). 
The four WCDA ranges at lower energy have normalized significance levels between $3.1\sigma$ and $7.2\sigma$, and the fit is to these four data values; values for the highest WCDA energy range and for KM2A have significance below $3\sigma$ and are plotted as upper limits (95\% confidence).
The best fit is for $\alpha=0.010\pm 0.001$ and the spectral index 
$\gamma = -0.53 \pm 0.12$. 

{\it Discussion.}---This is the first report of transient large-scale anisotropy in \rev{TeV} cosmic rays. We observe significant and consistent effects of this transient phenomenon across four $N_{\text{hit}}$ ranges, representing independent data sets, for median energies spanning from 0.7 to 3.1 TeV. When examining the relationship between maximum gradient magnitude \rev{$g$} and median energy $E$, we find that it follows a power-law form.
\rev{For} the median energy of 1 TeV, \rev{our result for $g$} corresponds to a dipole anisotropy of \rev{at least} 1\%, or alternatively, if the anisotropy pattern has a higher order, it can have a lesser magnitude;
see Supplemental Material for details.

\rev{Weak} dependence of transient anisotropy on energy is consistent with previous findings for solar-heliospheric effects in GeV-range CRs, including observations around the time of this ICME passage \citep{Munakata2022}, and is the reason why anisotropy effects can \rev{be} observed up to very high energies. 
Similarly, for modulation associated with the 11-year solar activity cycle, anisotropy effects have reported up to higher median energies (about 0.6 TeV) than effects on CR intensity \citep{munakata2010solar}.


Physically, the times with skymap gradients of strong normalized significance correspond to the end of  the sheath region between ICME and the preceding interplanetary shock, which arrived at Earth at 19:42 on November 3\footnote{http://isgi.unistra.fr}, and also to the start of the ICME passage, which lasted from November 4, 12:00 to November 5, 06:00 \citep{Munakata2022}.
In terms of formal significance, there were also strong peaks on November 2 and 3, which were after the CME eruption and before ICME arrival at Earth.

This new observation raises questions about the origin of \rev{transient TeV} CR anisotropy of $\sim$1\%. 
\rev{For comparison,} the Compton-Getting anisotropy stemming from Earth's orbital motion has an amplitude of $\approx$0.05\%, and the sidereal CR anisotropy from outside the heliosphere has an amplitude of only $\approx$0.1\%.
According to Liouville's theorem for a Hamiltonian system, which applies for a static and smooth magnetic field, the particle phase space density remains constant along the trajectory from a point and direction of observation back to the entry point into the heliosphere, 
so smooth, static magnetic fields should not deflect the sidereal pattern to generate an observed large-scale anisotropy gradient of $\sim$1\%. 

Therefore, the observed anisotropy must be attributed to magnetic fields that are not static or not smooth. In the region between the ICME and its leading shock, as well as within the ICME itself, the magnetic field \rev{$\mathbf{B}$} is in motion at the ICME speed 
\rev{$\mathbf{V}$.  A unidirectional magnetic field region of extent $L$ from the observer leads to a motional electric field that accelerates or decelerates particles depending on their direction, generating a dipole anisotropy of $\gamma qVBL/E$ for cosmic rays of charge $q$, energy $E$, and spectral index $\gamma$.  For measured values of $V$ and $B$ \citep{Munakata2022} and the very generous assumption of unidirectional $\mathbf{B}$ over $L=0.5$ AU, we estimate an anisotropy magnitude $<0.3$\% for 1 TeV protons, which is insufficient to explain the observed transient anisotropy.}

\rev{Here} we propose an explanation based on random magnetic fields, which induce stochastic scattering of cosmic ray directions. The interplanetary magnetic field exhibited its strongest magnitude and fluctuations in the sheath region between the shock and the ICME \citep{Munakata2022,Benella2025}. Notably, LHAASO recorded the most pronounced gradients when this sheath region had almost completely passed by Earth and therefore was concentrated in one half of the sky.
\rev{Note also that this explanation does not apply to particles at much lower energy, which can be confined within the ICME.  
At the time of interest, 65-GV ions observed by muon detectors exhibited unidirectional and bidirectional anisotropy of $>$1\% that was nearly rigidity-independent \citep{Munakata2022}, but was organized relative to the local magnetic field and thus involved a physically distinct mechanism.}

\begin{figure}[t]
\includegraphics[scale=0.28]{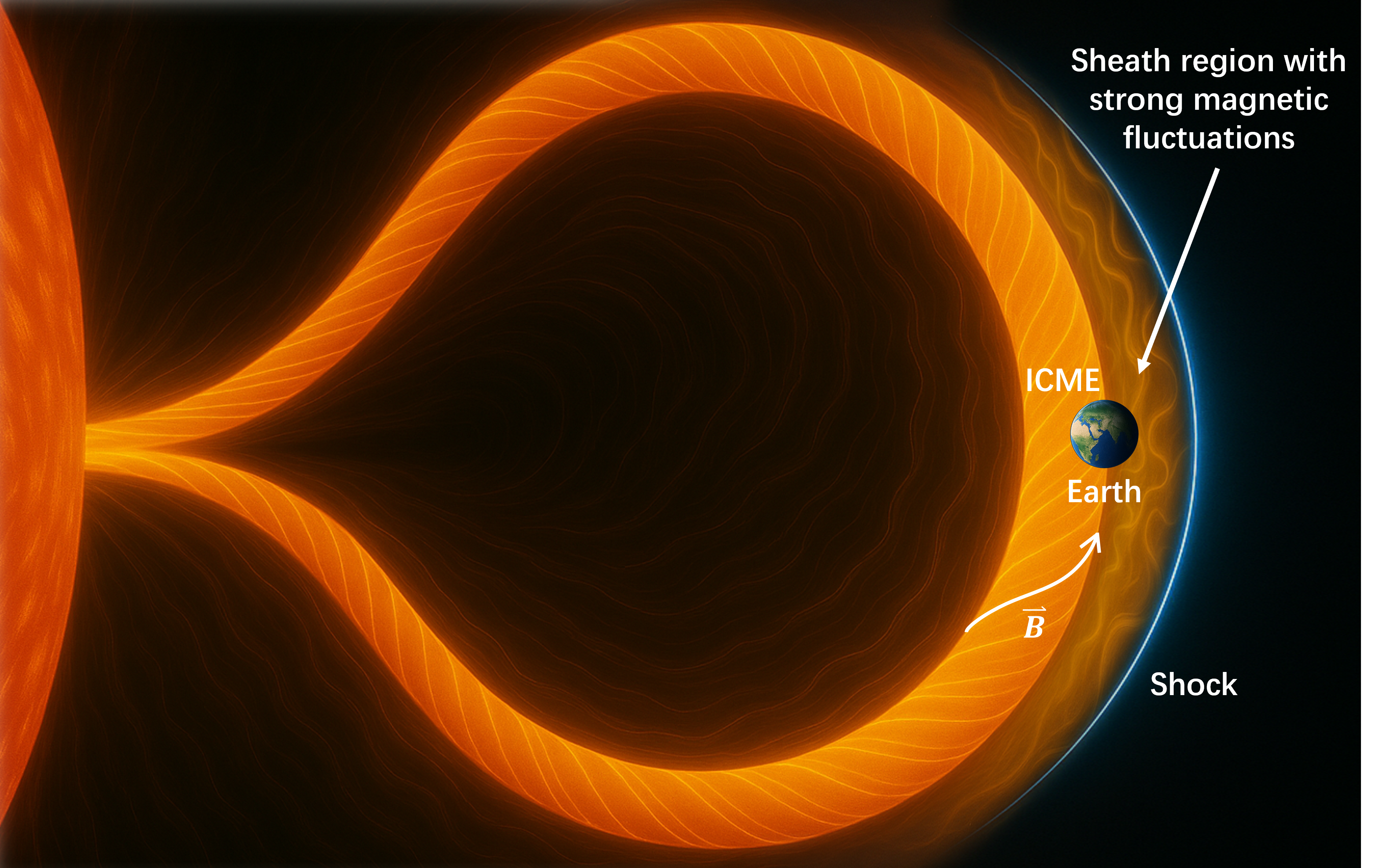}
\caption{Illustration of the magnetic flux rope of an ICME passing Earth on November 4, 2021, which was preceded by an interplanetary shock and a sheath region with intense magnetic fluctuations.
Transient anisotropy of \rev{TeV} CRs across LHAASO's FOV was strongest shortly before arrival of the leading edge of the ICME, with a lower flux in directions from the outer heliosphere.  
We attribute this to enhanced scattering of CRs along trajectories passing through the sheath region of enhanced magnetic turbulence.
}
    \label{fig:ICME}
\end{figure}

We hypothesize that transient anisotropy arises from varying scattering along different sky directions, as shown schematically in Figure \ref{fig:ICME}. Specifically, we anticipate the strongest scattering along paths intersecting more of the sheath region, predicting a pattern that is axisymmetric around the shock normal direction $\hat{n}$ \rev{(to the right in the Figure)}. 
The shock impacted Earth's magnetosphere with $\hat{n}=(-0.85, 0.52, 0.04)$ in GSE coordinates \citep{Regi22}, i.e., a nearly anti-Sunward direction at GSE longitude 149$^\circ$ and latitude 2$^\circ$. Our method is expected to register a minor gradient when the FOV includes an axis of symmetry, such as $\hat{n}$ or $-\hat{n}$, which roughly aligns with the pattern in Figure \ref{fig:WCDA_nhit30-40_GSE}.

Other air shower arrays worldwide can contribute skymaps of different sky regions at the same times, potentially constructing a global view of the anisotropy. Our new observation opens the door to studying interplanetary magnetic structures using air shower arrays worldwide, complementing existing {\it in situ} measurements and remote plasma density sensing by providing insights into magnetic fluctuations over extended interplanetary distances.

\begin{acknowledgments}

We would like to thank all staff members who work at the LHAASO site above 4400 meters above sea level year round to maintain the detector and keep the water recycling system, electricity power supply and other components of the experiment operating smoothly. We are grateful to Chengdu Management Committee of Tianfu New Area for the constant financial support for research with LHAASO data. We appreciate the computing and data service support provided by the National High Energy Physics Data Center for the data analysis in this paper. This research work is supported by the following grants: In Thailand, by the National Science and Technology Development Agency (NSTDA) and the National Research Council of Thailand (NRCT) under the High-Potential Research Team Grant Program (N42A650868) and from the NSRF via the Program Management Unit for Human Resources \& Institutional Development, Research and Innovation (B39G670013), and in China by the Project for Young Scientists in Basic Research of the Chinese Academy of Sciences No.YSBR-061, and by the National Natural Science Foundation of China, No.12273114, No.12321003, NO.12393851,
 No.12393852, No.12393853, No.12393854, No.12205314, No.12105301, No.12305120, No.12261160362, No.12105294, No.12375107,  No.12173039,  the Department of Science and Technology of Sichuan Province, China No.24NSFSC2319.
\end{acknowledgments}

\section{SUPPLEMENTAL MATERIAL}

\subsection{Data Selection}

\begin{figure}[h]
\includegraphics[scale=0.75]{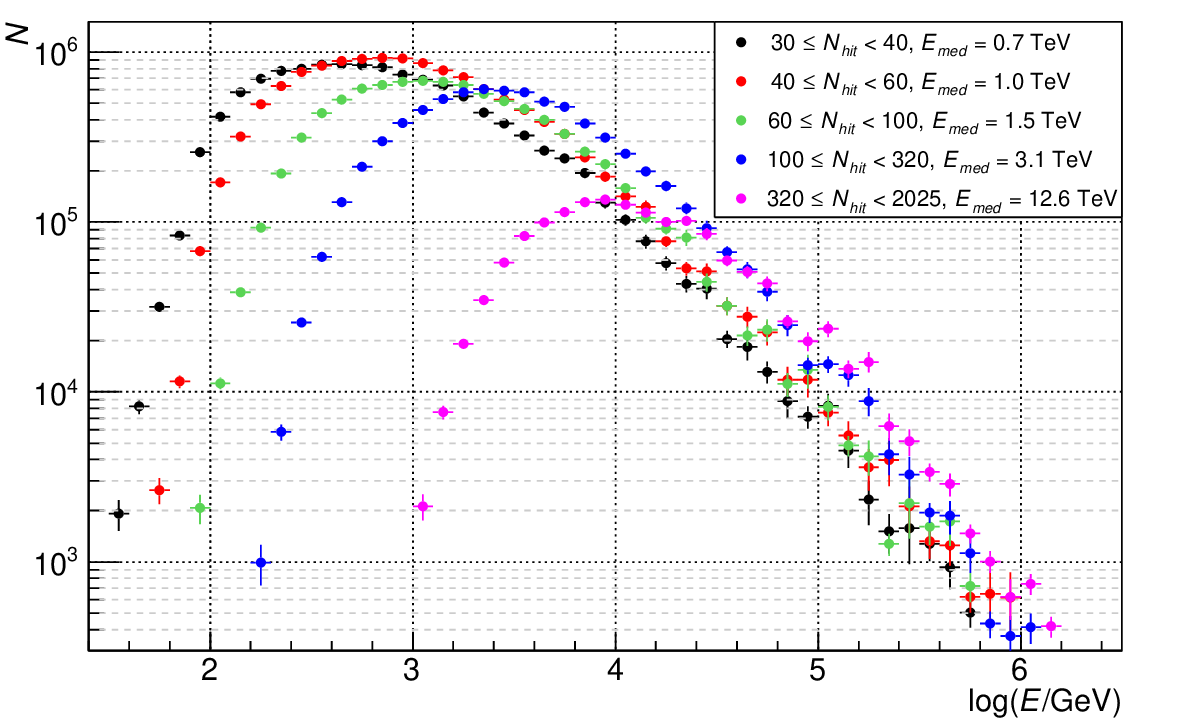}
\caption{Monte Carlo simulation of energy response for each WCDA $N_{hit}$ range that we use, with the median primary CR energy $E_{med}$ indicated in the legend.}
    \label{fig:WCDA_Edistribution}
\end{figure}

\begin{figure}[h]
\includegraphics[scale=0.5]{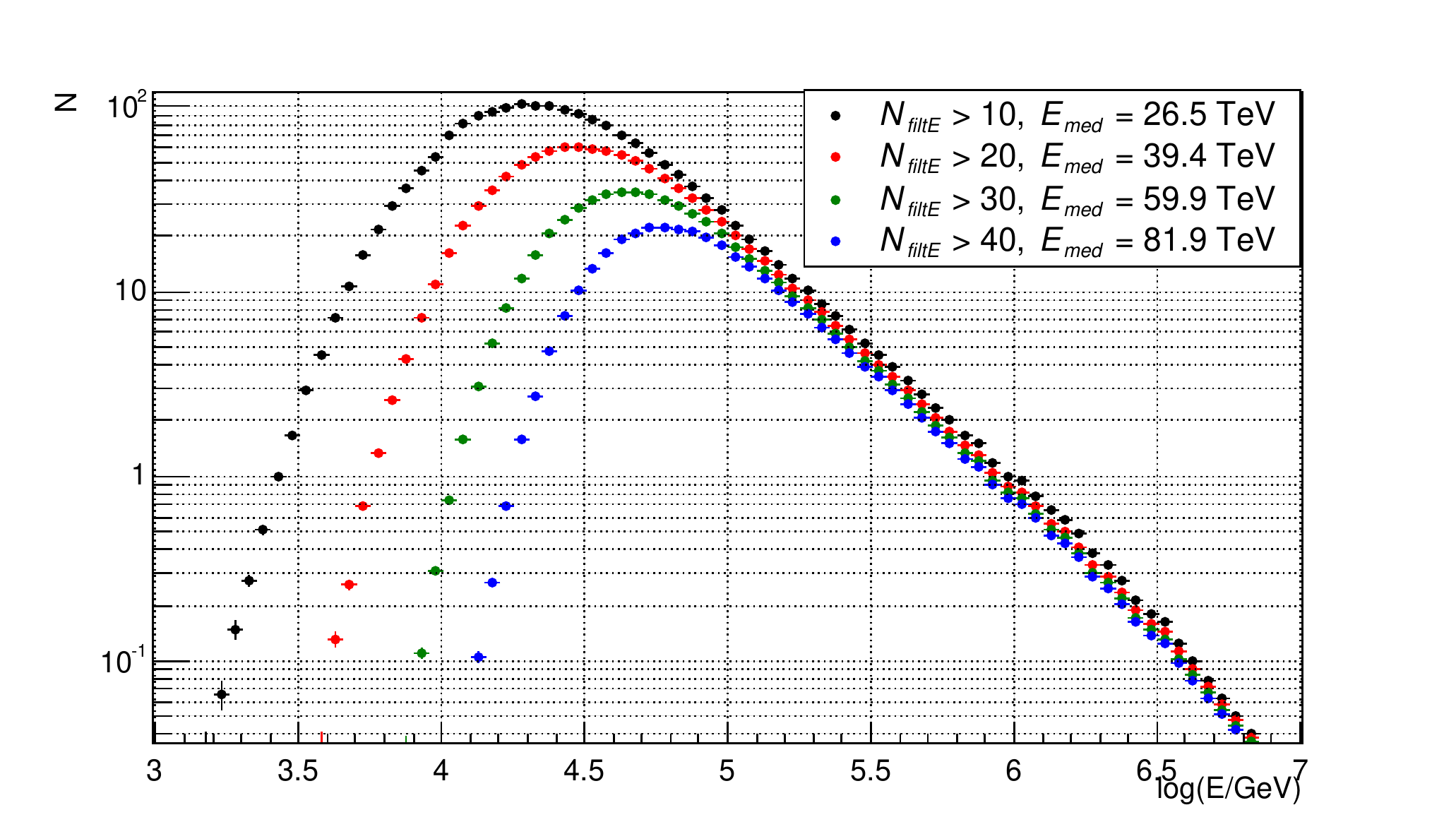}
\caption{Monte Carlo simulation of energy response for each KM2A $N_{filtE}$ range that we use, with the median primary CR energy $E_{med}$ indicated in the legend.}
    \label{fig:KM2A_Edistribution}
\end{figure}

Data were analyzed by selecting reconstructed shower events with zenith angles less than 45° for better statistical quality and to reduce biases related to the different relationship between primary CR energy and \(N_{\text{hit}}\) \rev{or \(N_{\text{filtE}}\)} at larger zenith angles.  For LHAASO-WCDA data, we selected hours with 100-120 million shower events.
\rev{The variable \( N_{hit} \), defined as the number of triggered detectors that record more than 0.5 photoelectrons within a 30 ns time window, is used for shower reconstruction and serves as a proxy for the energy of the primary cosmic ray \citep{LHAASO5}.
} Events were categorized into five energy bins for \( N_{hit} \) between these values: 30, 40, 60, 100, 320, and 2025. The median energies $E_{med}$ for these bins are 0.7, 1.0, 1.5, 3.1, and 12.6 TeV, respectively. \rev{For the LHAASO-KM2A data, we selected hourly intervals containing 7–10 million shower events. The parameter \( N_{filtE} \), defined as the number of triggered detectors within a –50 ns to 100 ns time window whose distances from the shower core, when projected onto the shower plane, are less than 200 m, was required to exceed 10, 20, 30, or 40.
These selections correspond to median energies of approximately 26.5, 39.4, 59.9, and 81.9 TeV, respectively.}
\rev{These median energy estimates are based on CORSIKA simulations \citep{heck1998corsika} of} primary cosmic rays, mainly including five components: protons, helium, CNO, MgAlSi, and iron. For WCDA, \rev{we use the poly-gonato model, in which} these components contribute 93.76\%, 5.53\%, 0.36\%, 0.10\%, and 0.06\% of the total, respectively, \rev{including} additional minor components\cite{Horandel03,LHAASO5}. For KM2A, only these five components are included, with corresponding fractions of 31.00\%, 23.46\%, 17.75\%, 15.16\%, and 12.63\% \cite{Gaisser12,aharonian_observation_2021}. A total of 1.5 million events are simulated for WCDA and 9.5 million for KM2A. We select events with zenith angles less than 45°, and plot their energy distributions (Figures \ref{fig:WCDA_Edistribution} and \ref{fig:KM2A_Edistribution}) based on $N_{hit}$ \rev{or $N_{filtE}$}.

\subsection{Details of the Analysis of Relative Intensity and Anisotropy\label{sec:SM-newmethod}}

In our analysis, for each hour of data, we create a skymap in the horizontal coordinate system, dividing it into segments of equal solid angle. There are eight values of the zenith angle \(\theta\), equally spaced in \(\cos\theta\) up to \(\theta = 45^\circ\), and sixteen values of the azimuth angle \(\phi\) (see Figure 2). This method allows us to generate 24 skymaps per day.

We define \(N_{i,t,\theta,\phi}\) as the count at zenith angle \(\theta\) and azimuth angle \(\phi\) for the \(i\)-th day and \(t\)-th hour. 
Note that CR showers attenuate with atmospheric depth, so the count depends strongly on $\theta$ and only weakly on $\phi$.
The relative (``zenith-normalized'') intensity of a cell at the time of interest (``on-source''), compared with the azimuthally averaged intensity \(\langle N_{i,t,\theta} \rangle\), is given by
\begin{align}
I^{on}_{i,t,\theta,\phi}=\frac{N_{i,t,\theta,\phi}}{\langle N_{i,t,\theta} \rangle}.
\end{align}
The background intensity for a given bin is estimated using the average relative intensity from the same \(t\)-th hour and \(\theta\) and \(\phi\) bin over the preceding two days and the subsequent two days, excluding the \(i\)-th day:
\begin{align}
I_{i,t,\theta,\phi}^{bkg} = \frac{\sum_{n=i-2}^{i+2} N_{n,t,\theta,\phi}}{\sum_{n=i-2}^{i+2} \langle N_{n,t,\theta} \rangle}. \quad (n \neq i)
\end{align}
Subsequently, the final relative intensity for each cell in the skymap is determined as:
\begin{align}
I_{i,j,\theta,\phi} = \frac{I_{i,j,\theta,\phi}^{on}}{I_{i,j,\theta,\phi}^{bkg}}.
\label{eq:I}
\end{align}
Our approach leverages the stability of known anisotropies, such as the sidereal anisotropy and the Compton-Getting effect due to Earth's orbital motion, which remain stable over several days in a given sky direction (i.e., for a given hour $t$). Thus our method naturally eliminates these influences.  

\begin{figure}[h]
\includegraphics[scale=0.6]{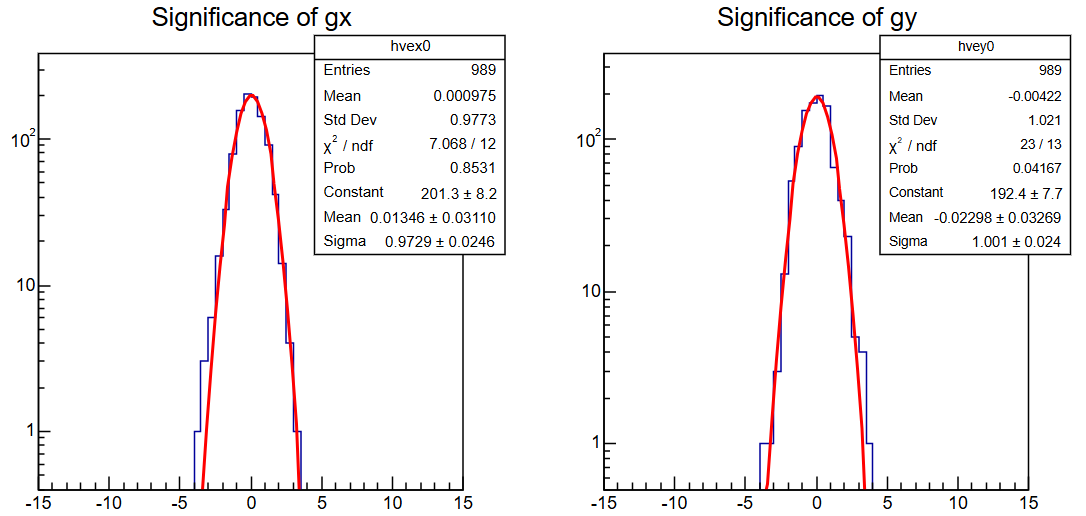}
\caption{Distributions of $S_x=g_x/\sigma_g$ and $S_y=g_y/\sigma_g$ for 45 days of quiet-time data, for WCDA data with $30\leq N_{hit}<40$, compared with Gaussian fits.}
    \label{fig:Sdist}
\end{figure}

The gradient $\mathbf{g}$ over a skymap is determined as described in the main text, and the formal fitting uncertainty of $g_x$ and $g_y$ for each hour is determined as part of the fitting procedure.  
As described in the main text, a multiplicative factor of $w$, which depends on energy and is constant in time, is applied so that the resulting uncertainty $\sigma_g$ accurately reflects the variance of $g_x$ and $g_y$ during quiet times.  Figure \ref{fig:Sdist} shows the resulting distributions of $S_x=g_x/\sigma_g$ and $S_y=g_y/\sigma_g$ for 45  days of quiet-time data.
\rev{Since $g_x$ and $g_y$ are well described by Gaussian distributions of width $\sigma_g$, the gradient magnitude $g$ is well described by the Rayleigh distribution 
\begin{align}
P(g) = \frac{g}{\sigma_{g}^2} e^{-g^2/(2\sigma_g^2)}.
\end{align}
}

\subsection{Comparison of Monthly Anisotropy Maps with Background Anisotropy \label{sec:SM-oldmethod}}

As a reality check on our analysis technique, we have confirmed that during quiet times without significant ICME effects, we can reproduce the known ``background'' pattern of sidereal diurnal anisotropy plus solar diurnal anisotropy.  
The sidereal anisotropy indicates an intensity pattern of GCRs in celestial coordinates before they enter our solar system (e.g., \citep{Bartoli18}). 
For this we use the measured long-term WCDA data in a given $N_{hit}$ range. 
We also consider the Compton-Getting (C-G) effect related to the motion of Earth toward a nearly isotropic distribution of cosmic rays in space \cite{ComptonGetting1935}, which increases the intensity of cosmic rays moving counter to Earth's orbital motion. Assuming a cosmic ray spectrum $\propto E^\Gamma$, we express the C-G anisotropy as 
\begin{equation} \label{C-G-Effect}
F_{C-G} (\theta) = (\Gamma-2)\frac{v}{c} \cos{\theta},
\end{equation}
where we use $\Gamma=-2.725$ for $E\sim1$ TeV based on \citep{Aguilar2015}, $v$ is the velocity of Earth's orbit around the Sun, $c$ is the velocity of light, and $\theta$ is the angle from the GSE $y$-direction. The conversion between GSE and celestial coordinates depends on the time of year, so these two ``background anisotropy'' components combine differently at different times of year.
We check whether our skymap analysis can properly detect this expected background for a given hour of the day (corresponding to a specific FOV in the sky) by combining all days in one month, choosing the duration of one month as an interval over which the combination of sidereal and C-G anisotropies does not change substantially.

\begin{figure}[!htb]
    \centering
    \includegraphics[width=0.9\textwidth]{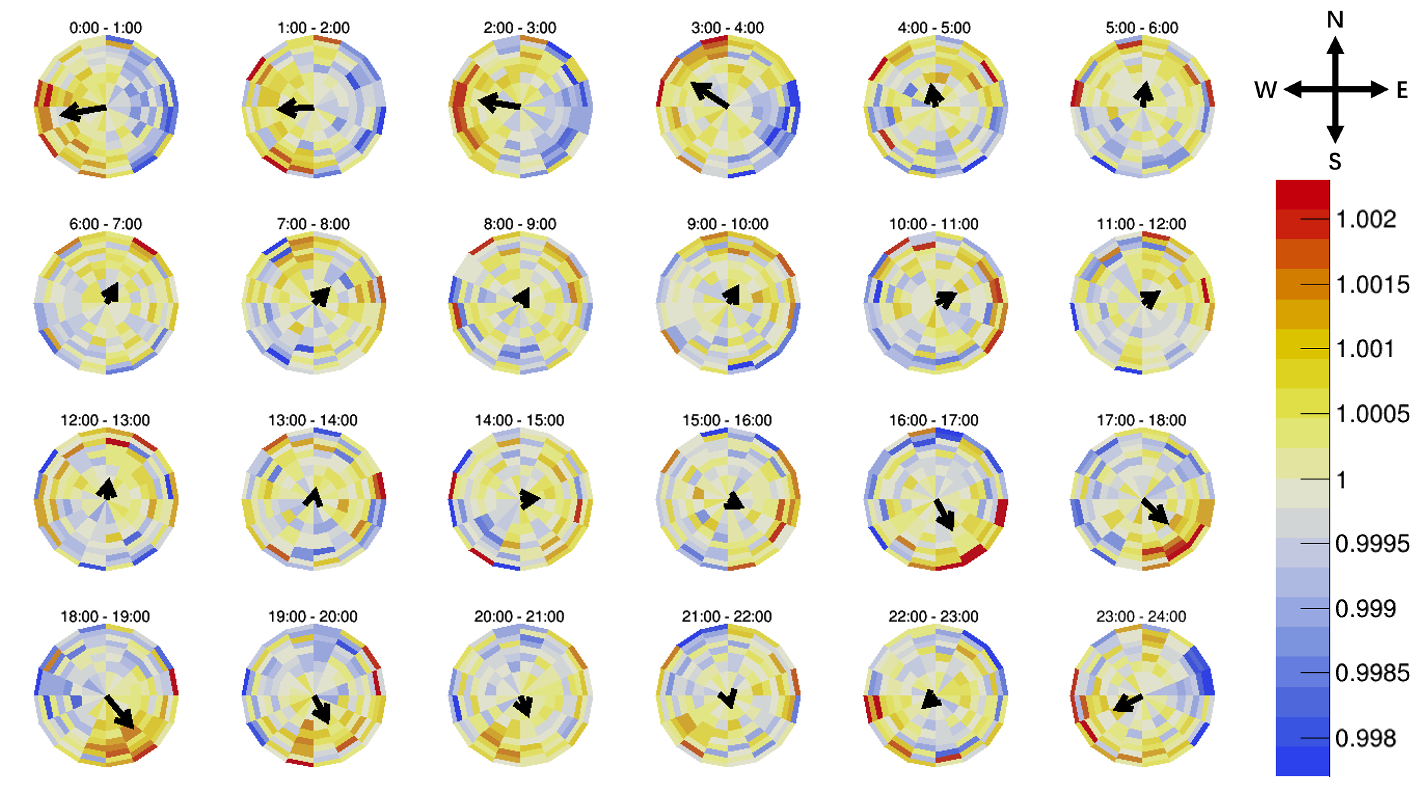}
    \caption{Zenith-normalized skymaps \rev{of CR intensity} for each hour of day \rev{averaged over} the quiet month of October 2021, for WCDA data with $60 \leq nhit < 100$. \rev{Black arrow represents the best-fit CR gradient vector, with arrow length proportional to gradient magnitude; the maximum magnitude in this Figure is 0.16\%.}
    }
    \label{fig:QuietTime}
\end{figure}

\begin{figure}[!htb]
    \centering
    \includegraphics[width=0.9\textwidth]{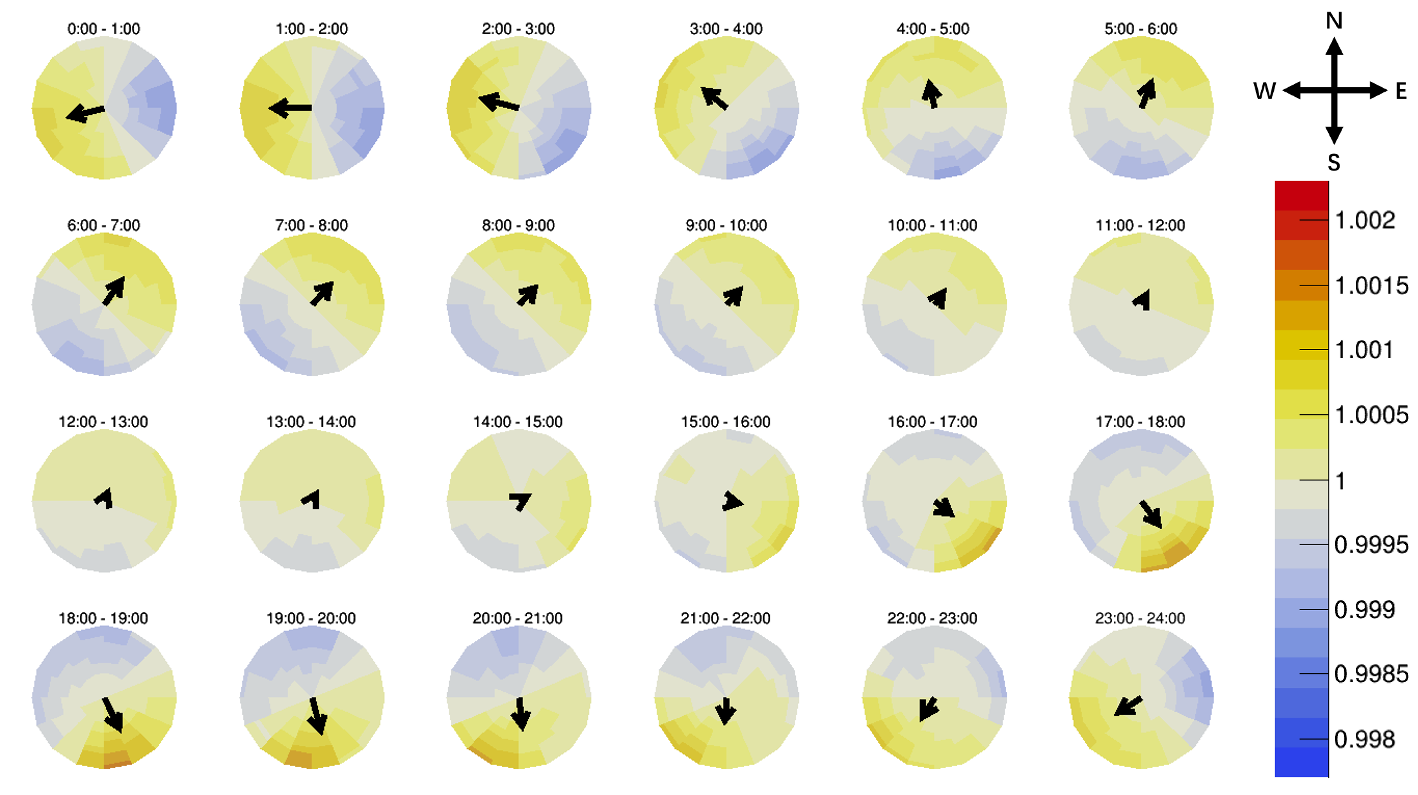}
    \caption{Like Figure \ref{fig:QuietTime}, but showing expected skymaps for the known background anisotropy during October 2021. 
    \rev{The maximum gradient magnitude is 0.14\%.}
    }
    \label{fig:Back-Com}
\end{figure}

For this purpose, we explored the zenith-normalized cosmic ray directional distribution. 
While the analysis method of the main text focuses on a differential comparison between the skymap for hour of day $t$ and those for hour $t$ of the preceding and following two days, here we consider a non-differential, zenith-normalized intensity skymap for hour $t$ of every day in a month,
and we divide by a monthly averaged skymap (averaged over all hours of all days) to remove instrumental non-uniformity (e.g., from the rectangular geometry of WCDA \cite{LHAASO5}, which favors some azimuthal directions over others).  
As an example, Figure \ref{fig:QuietTime} shows such skymaps for the quiet time of October 2021, when there were no significant ICMEs, for WCDA data with $60 \leq N_{hit} < 100$. 
The arrow on each map represents the gradient vector, with a length proportional to the gradient magnitude, according to the scale indicated under the color bar. 
The expected skymaps for the known background anisotropy (combined sidereal plus C-G anisotropy) are shown in Figure \ref{fig:Back-Com}.

\begin{figure}[!htb]
\includegraphics[width=\textwidth]{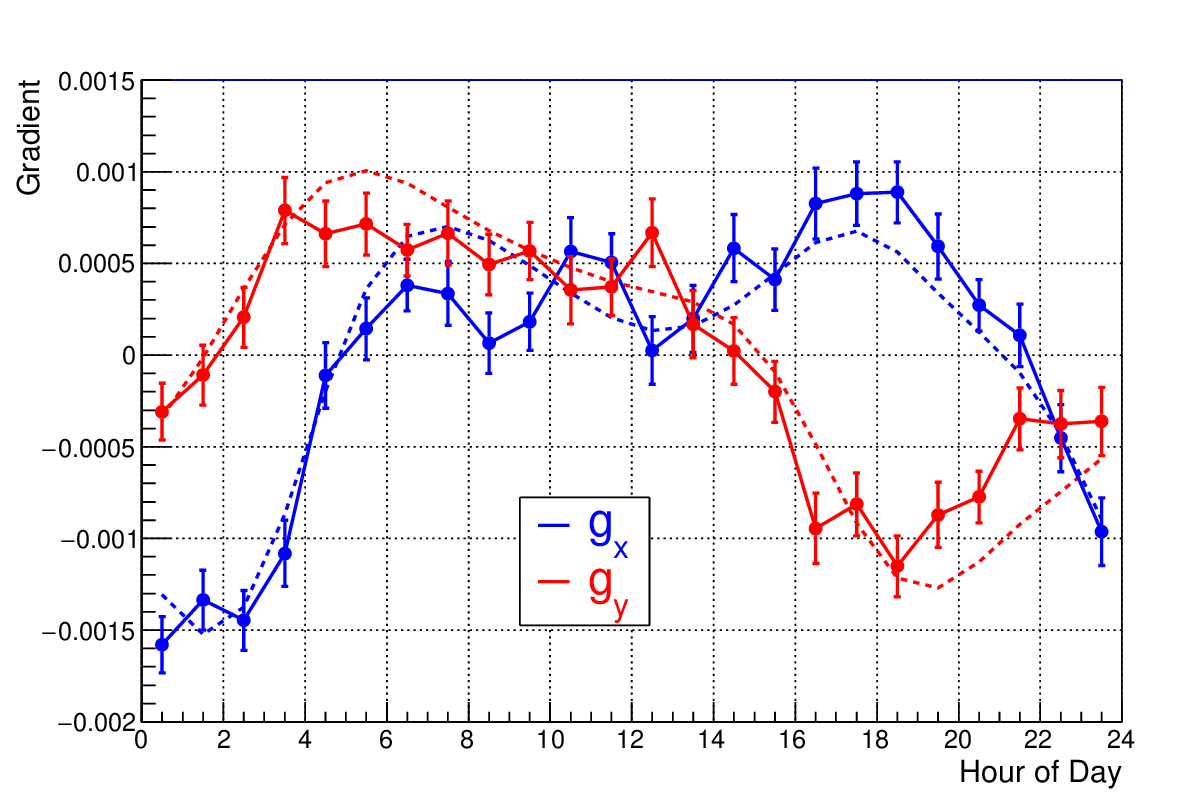}
\caption{
Skymap gradient components and magnitude for each hour of day, averaging over every day of October 2021, indicating consistency between the measured results \rev{(symbols with solid lines)} and expected background anisotropy effect \rev{(dashed curves)} during a quiet month.
}
    \label{fig:Grbe}
\end{figure}

The skymaps in Figures \ref{fig:QuietTime} and \ref{fig:Back-Com} appear to be consistent.
As a quantitative check, we also compute the skymap gradient $\mathbf{g}$ as described in the main text.
Figure \ref{fig:Grbe} compares the measured gradient components and magnitude (symbols \rev{with solid lines}) with the expected values for the background anisotropy (\rev{dashed} curves), indicating consistent results.
We also verified good consistency during other quiet months and for other energy ranges.  

As another, non-differential analysis method for examining transient anisotropy, a zenith-normalized skymap, also normalized to the monthly average, can be examined for each individual hour of data.  
We can calculate the gradient vector $\mathbf{g}$ and subtract out the gradient for the known background anisotropy. 
This analysis method can also detect the effect of the ICME on November 4, 2021.


\subsection{Inference about the Anisotropy Magnitude from the Gradient Magnitude}

\rev{When} interpreting our results, it is important to consider the limited FOV of our skymaps (Figure 2).  Any true anisotropy pattern will have at least one local minimum and one local maximum, and if the local zenith direction is near such an extreme point, the gradient should be suppressed. 
\rev{Therefore,} while observation of a strong gradient is evidence of a strong anisotropy pattern, observation of a weak gradient or no significant gradient near the local zenith direction does not necessarily imply that the overall anisotropy pattern is weak.
In our observations this may have occurred during 15:00-16:00 UT, when the gradient magnitude weakened as Earth rotated and the LHAASO FOV approached the anti-Sunward direction.  
After further rotation of Earth, during 16:00-17:00 the gradient was observed in a different direction.

\rev{Consider a pattern with ideal dipole anisotropy $\delta$, such that the relative intensity is $I=1+\delta\cos\alpha$, where $\alpha$ is the angle from the dipole axis.
If the FOV is centered along a sky direction with maximal gradient (i.e., along a plane perpendicular to the dipole axis), and we assume without loss of generality that the $x$-direction lies along the dipole axis, then our technique measures $(g_x,g_y)=(\delta,0)$ and $g=\delta$.  
However, if the FOV is oriented along another direction, then $g<\delta$. 
As a result, when we measure $g$ only over our field of view and do not know the orientation of the overall anisotropy pattern, we must infer that a dipole anisotropy pattern would have $\delta\geq g$.
In our case, for} a median energy of 1 TeV \rev{we measure $g\approx0.01$}, so for a dipole anisotropy pattern, this would imply an anisotropy of at least 1\%.

\rev{
Now consider a higher-order anisotropy pattern, and suppose that it has a maximum relative anisotropy of $\delta$ as above. 
In this case we could have a stronger gradient, such that $g>\delta$.  
Therefore, we conclude that our measurement of $g\approx0.01$ at median primary energy 1.0 TeV implies a dipole anisotropy 
of at least 1\%, or alternatively, a higher-order anisotropy that could have a lesser magnitude.}

\clearpage

\bibliographystyle{apsrev}
\bibliography{refs}

\end{document}